\documentclass[12]{iopart}
\usepackage{amsmathred, latexsym, amssymb, amscd,amsfonts, amsthm, graphicx,color,times,txfonts}
\usepackage{pictex}
\usepackage[english]{babel}

\newtheorem{theorem}{\bf Theorem}[section]

\begin{document}

\title[Asymptotically optimal quantum channel reversal]
{{\sf \bfseries Asymptotically optimal quantum channel reversal for qudit ensembles and multimode  Gaussian states}}
\author{ \sf \bfseries  Peter Bowles, M\u{a}d\u{a}lin Gu\c{t}\u{a} and Gerardo Adesso }
\address{School of Mathematical Sciences, The University of Nottingham,
University Park, NG7 2RD Nottingham, United Kingdom}

\date{}


\begin{abstract}
We investigate the problem of optimally reversing the action of an arbitrary quantum channel $C$ which acts independently on each component of an ensemble of $n$ identically prepared $d$-dimensional quantum systems. In the limit of large ensembles, we construct the optimal reversing channel $R_n^\star$ which has to be applied at the output ensemble state, to retrieve a smaller ensemble of $m$ systems prepared in the input state, with the highest possible rate $m/n$. The solution is found by mapping the problem into the optimal reversal of Gaussian channels on quantum-classical continuous variable systems, which is here solved as well. Our general results can be readily applied to improve the implementation of robust long-distance quantum communication. As an example, we investigate the optimal reversal rate of phase flip channels acting on a multi-qubit register.
\end{abstract}

\clearpage

\tableofcontents

\title[Asymptotically optimal quantum channel reversal]
\bigskip
\section{Introduction}



Quantum channels are completely positive, trace preserving maps which describe the state change of a quantum system  undergoing a noisy evolution. Operationally, a channel on a $d$-dimensional system
$C:M(\mathbb{C}^d) \longrightarrow  M(\mathbb{C}^d)$ can be implemented by first coupling the system with state
$\rho$ to the ``environment'' with initial state  $\rho_{E}$, letting the two evolve together as a closed system with unitary operator $U$, and subsequently tracing out the environmental degrees of freedom \cite{NC,petruccione}
\begin{equation}
\label{eq.channel.coupling}
C:\rho \longmapsto  {\rm Tr}_{E} ( U (\rho\otimes \rho_{E} )U^{*}).
\end{equation}
 Due to the last step, the dynamics described by a quantum channel is typically irreversible, a basic fact which is related to several no-go theorems in quantum information \cite{WZ,KW,J,SH}. In fact, a channel
is reversible \emph{on all states} if and only if it is unitary i.e.
$C(\rho) = U\rho U^{*}$.

However, the reversibility problem has non-trivial solutions if the channel is required to be reversible on a given \emph{family of states} \cite{SH}.
From a mathematical perspective, this scenario is captured by the the statistical concepts of quantum \emph{sufficiency} and \emph{equivalence of statistical models}. In the latter, two families of states
$$
\mathcal{Q} :=\{\rho_{\theta} :\theta\in \Theta\}   \quad \text{and}\quad
\mathcal{R}:= \{\sigma_{\theta} :\theta\in \Theta\}
$$
are said to be statistically equivalent if there exist two channels
$T$ and $S$ such that $\sigma_{\theta}= T(\rho_{\theta})$ and $\rho_{\theta} = S(\sigma_{\theta})$ for all $\theta$. In other words the channel $T$ is reversible on $\mathcal{Q}$, with an inverse $S$. Sufficiency is a special case of this, where
$T(\rho)$ is the restriction of the state $\rho$ to a subalgebra of observables, which is then called sufficient with respect to
$\mathcal{Q}$. A general mathematical characterization of equivalence and sufficiency has been given in \cite{JP}, where a quantum analogue of the classical factorization theorem for sufficient statistics has been established.

\begin{figure}[t]
\begin{center}
 \includegraphics[width=8.5cm]{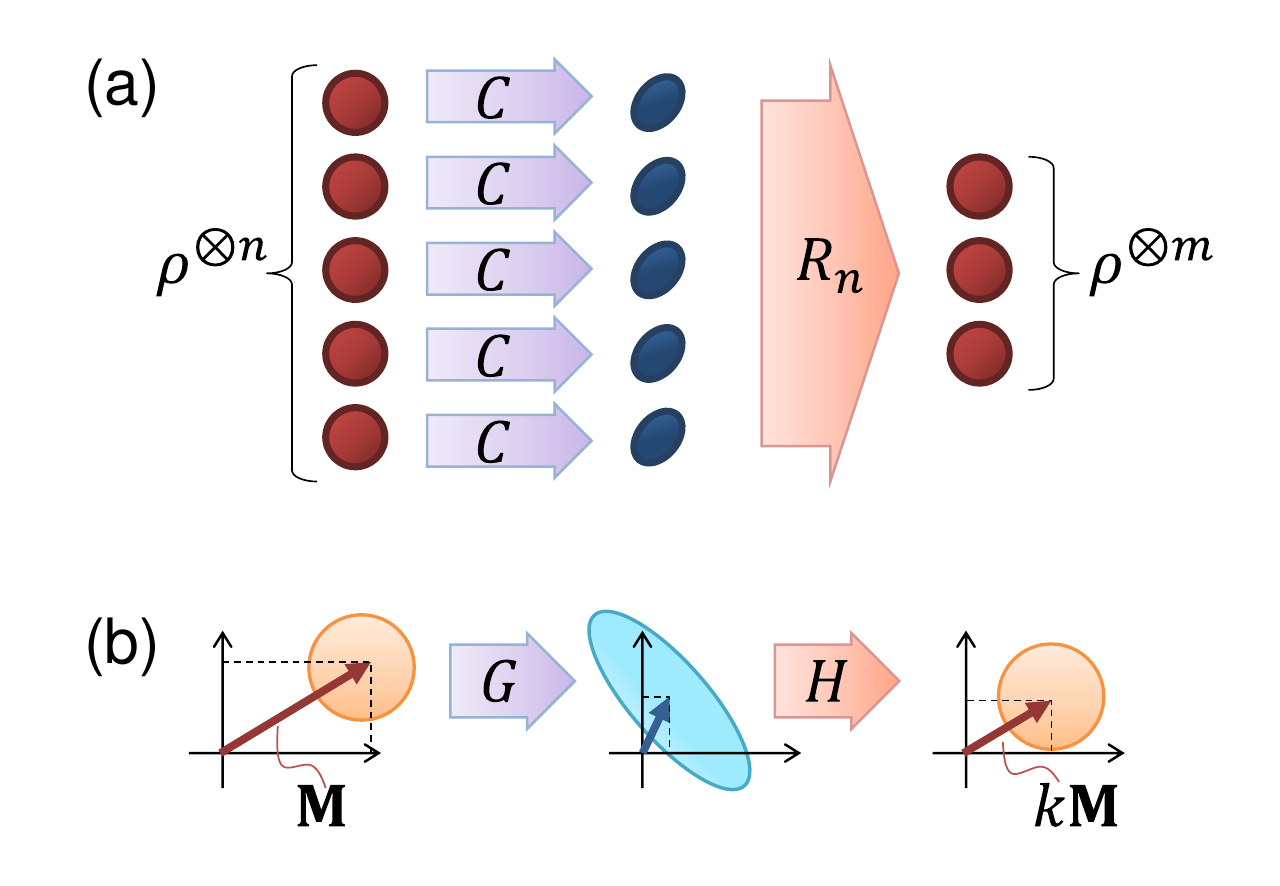}
\caption{(Color online) Graphical depiction of the problem addressed in this paper. (a) Reversal of arbitrary channels $C$ acting independently on a register of $n$ qudits, each initialized in the state $\rho$. The optimal reversing channel $R_n$ allows to recover $m$ qudits exactly in the state $\rho$, with the maximum possible output/input rate $m/n$. (b) The problem can be recast into the reversal of a Gaussian channel $G$ acting on a set of Gaussian states with unknown mean ${\bf M}$. The optimal reversing channel $H$ restores the initial known covariance and maximizes the output/input displacement ratio $k$. More details are provided in the diagram (\ref{cd}) and in the accompanying text.}
\label{figproblem}
\end{center}
\end{figure}

In this paper we consider a different but related channel reversal problem, see Fig.~\ref{figproblem}. Consider an ensemble consisting of  $n\gg 1$  independent and identically prepared quantum systems which undergo separate noisy evolutions described by the channel $C$.
The question is whether the original state can be distilled from the noisy output by means of a reversing channel, albeit at the expense of reducing the size of the ensemble due to the loss of information. More precisely, we would like to find the maximum rate $\Gamma=  m /n$ for which there exist reversing channels $R_n:M(\mathbb{C}^d)^{\otimes n}\longmapsto M(\mathbb{C}^d)^{\otimes m}$ such that  the  following holds
\begin{equation}
\rho^{\otimes n}\stackrel{C^{\otimes n}}{\longmapsto}C(\rho)^{\otimes n}\stackrel{R_n}{\longmapsto}\rho^{\otimes m}
\label{prob}
\end{equation}
asymptotically with $n$ (see section \ref{c3} for the mathematical formulation).  This and other related questions have been considered in \cite{KW,BGA} for the specific case of a qubit depolarizing channel, but to our knowledge the case of a general qudit channel has not been investigated elsewhere. The answer to our question is directly relevant to applications in quantum memories \cite{atomicmemories,polzik}, distributed quantum computation \cite{seralefibers}, quantum key distribution \cite{manyqkd},  and long-distance quantum communication  \cite{quantuminternet}. Given a register of many qudits prepared off-line, in order  to securely store it or transmit it over lossy media,  one needs to incorporate modules that perform state purification or  counteract decoherence effects, e.g.
{\it quantum repeaters} \cite{repeaters}.

%


The key to solving the problem lies in the quantum extension of a fundamental statistical tool known as local asymptotic normality (LAN) \cite{LC,Young&Smith}.
The quantum version of LAN \cite{GK, GJK, GK2, GJ, GKJ} shows that the joint state of an ensemble of identically prepared systems can be approximated by a Gaussian state of a continuous variable system. More precisely, we parametrize the states in a neighbourhood of size $n^{-1/2+\epsilon}$ of
$\rho$ as
$
 \rho_{\textbf{M}/\sqrt{n}}
$
where ${\bf M}$ is a $(d^{2}-1)$-dimensional vector of expectations such that $\rho_{\textbf{0}} \equiv \rho$; then there exists a channel $T_{n}$ which maps the collective state $\rho_{\textbf{M}/\sqrt{n}}^{\otimes n}$ into a Gaussian state $\Phi({\bf M}, {\bf V})$ of a quantum-classical continuous variable system, with mean ${\bf M}$ and known fixed covariance ${\bf V}$ which depends only on $\rho$. Conversely, given the Gaussian state, there exists a channel $S_{n}$ which converts it into the multiple-qudits state with asymptotically vanishing norm-one error, without the knowledge of the parameter ${\bf M}$.


The schematic of the complete problem considered in this paper and its resolution using LAN is shown below (see Fig.~\ref{figproblem} for a visual summary):
\begin{equation}\label{cd}
\begin{CD}
\rho^{\otimes n}_{(\textbf{M},\boldsymbol{\lambda})} @>C^{\otimes n}>> \rho^{\otimes n}_{(\textbf{M}',\boldsymbol{\lambda}')}@>R_n>>\rho^{\otimes m}_{(\textbf{M},\boldsymbol{\lambda})}           \\
   @V{T_n}VV @V{T_n}VV   @AA {S_m} A      \\
   \Phi(\textbf{M},{\bf V})
   @> G >>\Phi(\textbf{M}',{\bf V}')@>H>> \Phi(k\textbf{M},{\bf V})  \\
   \end{CD}\end{equation}

Using the LAN correspondence between the multiple-qudits states and the Gaussian states, we demonstrate specifically the following results:
\begin{itemize}
\item
The action of $C^{\otimes n}$ on the qudits can be mimicked by the action of a Gaussian channel on the Gaussian states, i.e. the rightmost square loop of the diagram  \eqref{cd} is (asymptotically) commutative (cf. Theorem \ref{teoG});

\item
The optimal qudit reversal problem for the channel $C$ is effectively recast into a much handier Gaussian one, of reversal of the Gaussian channel $G$ \cite{EW,H,HW,oldserafozzi,newsimon} where the length of the final displacement vector is allowed to be a fraction $k$ of the initial length as illustrated in the bottom row of the diagram \eqref{cd}; the problem is to find the maximum possible $k$ and the corresponding reversing channel $H$ (cf. Theorem \ref{teoqudits});

\item
The Gaussian reversal problem has an explicit solution $(k^{\star},H^{\star})$ as described in Theorem \ref{teogauss}.
The optimal reversal of a general Gaussian channel is in its own right an important problem for continuous variable quantum communication and key distribution \cite{brareview,pirlandolareview}, which we solve in a specific case of input states associated with qudit ensembles.

\item
Putting together the above findings we prove that the concatenation $R_{n}:=S_m\circ H\circ T_n$ realizes the {\it optimal} strategy for reversal of the channel $C^{\otimes n}$ applied to $n$ independent and identically prepared qudits, and that the maximal qudit rate $m/n$ in the top row of the diagram (\ref{cd}) is equal to $(k^\star)^2$, where $k^{\star}$ is the maximum possible value of $k$ for which
the Gaussian reversal) is possible.
\end{itemize}

%


The paper is organized as follows. In section \ref{c2} we find the optimal procedure and rate for reversing the action of general Gaussian channels on a family of Gaussian states with fixed covariance and unknown mean. Section \ref{LF} contains a short review of LAN motivated by the quantum Central Limit Theorem. In section \ref{c3} we solve the problem of finding the optimal rate and procedure for the reversal of a qudit channel by reducing it to the Gaussian problem via LAN. In section \ref{c4} we illustrate the general results with a specific application to reversing phase flip channels acting on a multiqubit register, a common source of errors arising in quantum computation \cite{NC}. We draw our concluding remarks in section \ref{secconcl}. Some technical proofs are deferred to Appendixes.

%
\section{Optimal reversal of a Gaussian channel}
\label{c2}

In order to solve the qudit channel reversal problem, it is necessary to obtain the solution to the corresponding Gaussian problem. We begin with a short review of Gaussian states and channels, referring the reader to \cite{eisertplenio,brareview,ourreview,pirlandolareview} for more comprehensive accounts. 
\subsection{Gaussian states and channels}
\label{GF}
Let $\textbf{Z}=(\textbf{A},\textbf{B})^{T}$ be the (column vector of) coordinates of a continuous variable system where the components
$\textbf{A}^{T}=(Q_1,P_1,\dots,Q_{K},P_{K})$ are the canonical observables of $K$ modes, and
$\textbf{B}^{T}=(B_1,\dots,B_{C})$ is a classical $\mathbb{R}^{C}$-valued random variable.
In other words, the canonical variables satisfy the commutation relations $[Z_k,Z_l]=i \Omega_{kl}\mathbf{1}$ where
$\Omega$ is the block-diagonal symplectic form
$$
\Omega = \text{Diag}\left(\sigma,\dots,\sigma, \mathbf{0}_{C}\right), \qquad
\text{with}\qquad \sigma=
\left(
\begin{array}{cc}
 0 & 1 \\
-1 & 0
\end{array}
\right).
$$
Throughout, we consider that the quantum variables are represented in the standard fashion on the multi-mode Hilbert
space $\mathcal{H}^{\otimes K}:= L^{2}(\mathbb{R})^{\otimes K}$ and the classical variables are realized as coordinate multiplication operators on $\mathcal{H}^{\otimes C}$ so that the full Hilbert space is $\mathcal{H}^{\otimes (K+C)}$.
Any state $\Phi$ of the continuous variable system is completely determined by its \emph{characteristic function}
$$
\chi_{\Phi}(z):=\Phi\left(e^{iz^T\textbf{Z}} \right)
$$
where $\Phi(X)$ denotes the expectation of $X$ with respect to $\Phi$. In particular, for every mean
$\textbf{M}=\Phi(\textbf{Z})$ and covariance matrix ${\bf V}$ with elements
$$
V_{kl}=\Phi\left( Z_{k}\ast Z_{l}\right)-\Phi(Z_k)\Phi(Z_l), \qquad Z_{k}\ast Z_{l} := \frac{Z_k Z_l+Z_lZ_k}{2},
$$
there is a unique Gaussian state $\Phi(\textbf{M},\textbf{V})$ with characteristic function
\begin{equation}\label{eq.ch.fct.gaussian}
\chi_{\textbf{M},\textbf{V}}(z)=e^{iz^T\textbf{M}-\frac{1}{2}z^T \textbf{V} z},
\end{equation}
provided that $\textbf{V}$ satisfies the uncertainty principle $\textbf{V}\geq-\frac{i}{2}\Omega$. For later use, we
will denote by $\phi(\textbf{M},\textbf{V}) \in \mathcal{T}_{1}(\mathcal{H}^{K}))\otimes L^{1}(\mathbb{R}^{C})$ the density matrix of the state $\Phi(\textbf{M},\textbf{V})$.

A Gaussian quantum channel is a channel as in \eqref{eq.channel.coupling}, where the environment is a bosonic continuous variable system whose initial state is Gaussian, and the unitary $U$ is determined by a quadratic Hamiltonian in the system and environment coordinates \cite{EW,H,HW,oldserafozzi,newsimon}. In the Heisenberg picture, the action of a Gaussian channel $G$ is
\begin{eqnarray}
G^{*}&:&
\mathcal{B}(\mathcal{H}^{\otimes K})\otimes L^{\infty}(\mathbb{R}^{\otimes C})
\to
\mathcal{B}(\mathcal{H}^{\otimes K})\otimes L^{\infty}(\mathbb{R}^{\otimes C}) \nonumber \\
G^{*}&:& W_{z}\longmapsto W_{{\bf X}_G z }e^{-\frac{1}{2}z^T{\bf Y}_Gz}
\label{gch}
\end{eqnarray}
where  $W_z:=e^{iz^T\textbf{Z}}$ are the ``Weyl operators'',  ${\bf X}_G$, ${\bf Y}_G$ are  real matrices of dimension
$2K+C$, with ${\bf Y}_{G}$ positive and satisfying the matrix inequality
$$
{\bf Y}_G\geq  i({\bf X}_G^T\Omega {\bf X}_G-\Omega).
$$
In particular, from \eqref{eq.ch.fct.gaussian} and \eqref{gch} we find that when $G$ acts on the state $\Phi(\textbf{M},\textbf{V})$, it produces a new Gaussian state
$\Phi(\textbf{M}',\textbf{V}')$ with mean $\textbf{M}^{\prime}={\bf X}_G^T \textbf{M}$ and covariance $\textbf{V}^{\prime}={\bf X}_G^T\textbf{V}{\bf X}_G+{\bf Y}_G$.
While the first term in $\textbf{V}^{\prime}$ describes the change in variance due to the linear transformation $X_{G}$, the second term comes from the covariance of the ancillary ``environment'' used to realize $G$. For more details on Gaussian channels and their classification we refer to \cite{holevonew}.

\subsection{Optimal Gaussian channel reversal}

Our problem can be stated as follows: given a Gaussian channel $G$,  and a family of Gaussian states
\begin{equation}\label{eq.gaussian.shift}
\mathcal{G}_{V}:= \{ \Phi(\textbf{M},\textbf{V}) :\textbf{M} \in \mathbb{R}^{2K+C} \}
\end{equation}
with fixed covariance matrix ${\bf V}$ and unknown mean $\textbf{M}$, we would like to find the maximum value $k^{\star}$ of a constant $k$ for which there exists a (not necessarily Gaussian) channel $H$ such that for all $\textbf{M} \in \mathbb{R}^{2K+C}$ the following holds [see Fig.~\ref{figproblem}(b)]
\begin{equation}
H\circ G\left(\phi(\textbf{M},\textbf{V})\right)=\phi(k \textbf{M},\textbf{V}).
\label{minmax}
\end{equation}

A special case of this problem, for single-mode attenuation and amplification channels, has been considered in \cite{BGA}. Our first main result is to find $k^{\star}$ and the optimal channel for the general set-up.
\begin{theorem}\label{teogauss}
Let $G$ be the Gaussian channel \eqref{gch}, and assume that ${\bf X}_{G}$ is an invertible matrix.
The largest value of $k$ for which $G$ can be reversed on the  family of states \eqref{eq.gaussian.shift} is
\begin{equation}
k^{\star}=\left[\lambda_{\max}\left(\boldsymbol{\varsigma}^{-\frac12} \boldsymbol{\varrho} \boldsymbol{\varsigma}^{-\frac12}\right)\right]^{-\frac12}
\label{maxk}
\end{equation}
where $\lambda_{\max}({\bf A})$ denotes the maximum eigenvalue of ${\bf A}$, and
$$
\boldsymbol{\varrho}=({\bf X}_G^{-1})^T \left({\bf X}_G^T \mathbf{V} {\bf X}_G + {\bf Y}_G+\frac{i}{2}\Omega\right){\bf X}_G^{-1}, \quad
\boldsymbol{\varsigma}=\mathbf{V}+\frac{i}{2} \Omega,
$$
are positive matrices.

Equivalently, $k^{\star}$ can be expressed in terms of the  `max-relative entropy'  \cite{NIL} as
$$
\log_2 ({k^{\star}}^{-2}) = D_{\max}(\boldsymbol{\varrho}\|\boldsymbol{\varsigma}):=\log_2\left(\min\big\{k^{-2}:\boldsymbol{\varrho}\leq k^{-2}\boldsymbol{\varsigma}\big\}\right).
$$
For every $k\leq k^{\star}$ there exists a reversing Gaussian channel $H$, of the form \eqref{gch}, with
\begin{equation}
\begin{split}
{\bf X}_H&=k{\bf X}_G^{-1}\,,\\
{\bf Y}_H&=\mathbf{V}-k^2({\bf X}_G^{-1})^T({\bf  X}_G^T\mathbf{V}{\bf X}_G+{\bf Y}_G){\bf X}_G^{-1}\,.
\end{split}
\label{XHYH}
\end{equation}

\end{theorem}

{\it Proof.}
Following a standard argument \cite{O,BGA}, it can be shown that without loss of generality we can restrict our attention to a certain class of \emph{displacement covariant}  reversing channels of the form
$$
T^{*}(W_z)=f(z)W_{{\bf X}z} ,
$$
where ${\bf X} = k{\bf X}_{G}^{-1}$ and $f(z)=\text{Tr}(\tau W_z)$ 
is the characteristic function of some ancillary state $\tau$ with zero mean. A displacement covariant channel $H$ satisfies \eqref{minmax} if and only if
$$
H\circ G\left( \phi(\textbf{0},\textbf{V} )\right)  = \phi(\textbf{0},\textbf{V} ).
$$
In terms of the characteristic functions, this means
\begin{align*}
\chi_{\Phi}(z)&=\text{tr}\big(e^{iz^T\textbf{Z}}H(\phi')\big) \nonumber\\
&=\text{tr}\big(H(e^{iz^T\textbf{Z}})\phi'\big)=\text{Tr}(\tau W_z)\text{Tr}\big(\phi'e^{i({\bf X}_H z)^T\textbf{Z}}\big)\nonumber\\
&=f(z)e^{-\frac{1}{2}({\bf X}_H z)^T{\bf V}'({\bf X}_H z)}.
\end{align*}
Since $\chi_{\Phi}$ is a Gaussian characteristic functions, $f(z)=\text{tr}(\tau W_z)$ must also be a Gaussian characteristic function. Hence $H$ is a Gaussian channel.

Now, from (\ref{gch}) we see Gaussian channels are in one-to-one correspondence with two matrices: the linear transformation on the means and the covariance (noise) matrix. The Gaussian channel $G$ and the reversal $H$ are therefore completely characterized by $G\Leftrightarrow({\bf X}_G,{\bf Y}_G)$ and $H\Leftrightarrow({\bf X}_H,{\bf Y}_H)$. For an initial state $\Phi(\textbf{M},\textbf{V})$, we have that the first and second moments are mapped by $(H\circ G) $ as
$$
(\textbf{M},\textbf{V})\mapsto\big({\bf X}_H^T{\bf X}_G^T\textbf{M}, \,{\bf X}_{H}^T({\bf X}_G^T\textbf{V}{\bf X}_G+{\bf Y}_G){\bf X}_{H}+{\bf Y}_H\big)
$$
By equating this with the target state moments $(k\textbf{M},{\bf V})$, we obtain immediately ${\bf X}_H=k{\bf X}_G^{-1}$. Now, denoting $\textbf{V}'={\bf X}_G^T\textbf{V}{\bf X}_G+{\bf Y}_G$, we have that the reverse outputs a covariance
\begin{equation}
\textbf{V}={\bf X}_H^T \textbf{V}' {\bf X}_H+{\bf Y}_H=k^2 ({\bf X}_G^{-1})^T \textbf{V}' {\bf X}_G^{-1}+{\bf Y}_H\,.
\label{one}
\end{equation}
Positivity of the quantum channel states that
\begin{equation}
{\bf Y}_H+\frac{i}{2}\Omega-\frac{i}{2}k^2 ({\bf X}_G^{-1})^T\Omega {\bf X}_G^{-1}\geq0
\label{two}
\end{equation}
We would like to find the maximum $k$ for which there exists some ${\bf Y}_H$ which satisfies this inequality.
Rearranging (\ref{one}) as
$$
{\bf Y}_H=\textbf{V}-k^2 ({\bf X}_G^{-1})^T\textbf{V}' {\bf X}_G^{-1}
$$
then substituting in (\ref{two}) gives
\begin{equation}
\textbf{V}+\frac{i}{2}\Omega\geq k^2 ({\bf X}_G^{-1})^T (\textbf{V}'+\frac{i}{2}\Omega) {\bf X}_G^{-1}
\label{cs}
\end{equation}
This gives us a necessary and sufficient condition for a Gaussian channel $G$ to be reversible up to a factor $k$ in the displacement. The optimal reversing channel $H^\star$ is the one for which $k$ takes its maximum possible value $k^{\star}$. 
To find $k^{\star}$, we can recast (\ref{cs}) in terms of a max-relative entropy and use the results of \cite{NIL} to obtain Eq.~(\ref{maxk}).
\hfill $\square$

This result shows that any Gaussian channel $G$ acting on Gaussian states with given covariance, can be reversed up to a constant factor $k$ by means of another {\it Gaussian} channel $H$, whose construction is provided explicitly. The link between the threshold value $k^{\star}$ and the max-relative entropy \cite{NIL} is intriguing, as it reveals that $k^{\star}$ is operationally related to the optimal Bayesian error probability in determining which covariance matrix our system is mapped into by the channel $G$ \cite{mosonyidatta}.

\section{Quick review of Local Asymptotic Normality}
\label{LF}
The main tool used to solve the channel reversal problem on $n$-qudit ensembles in the next section of this paper is that of quantum local asymptotic normality (LAN) which was developed in \cite{GK, GJK, GK2, GJ, GKJ} as an extension of a key concept from classical asymptotic statistics \cite{LC}. For reader's convenience we give here an intuitive explanation of LAN based on the quantum Central Limit Theorem.

In general terms,  \emph{classical} LAN means that given data consisting of $n$ samples from a probability distribution $\mathbb{P}_{\theta}$ with unknown parameter $\theta\in \mathbb{R}^{k}$, there exist classical channels (randomizations) which map the data into a \emph{single} sample from a Gaussian distribution whose mean is (locally) equal to $\theta$, and whose variance is the inverse Fisher information matrix $I(\theta)^{-1}$. A consequence of this is the fact that the maximum likelihood estimator $\widehat{\theta}_{n}$ is asymptotically normal (Gaussian) with asymptotic variance $I(\theta)^{-1}$, i.e. the following convergence in distribution holds
$$
\sqrt{n} (\widehat{\theta}_{n}-\theta) \overset{\mathcal{L}}{\longrightarrow} N(0, I(\theta)^{-1})
$$
and therefore saturates the Cram\'{e}r-Rao bound \cite{Young&Smith}.

\emph{Quantum} LAN means that the joint quantum state of independent, identically prepared  finite-dimensional systems can be approximated in a strong sense by a quantum-classical Gaussian state of fixed variance, whose mean encodes the information about the parameters of the original state. In this way, a number of asymptotic problems can be reformulated in terms of Gaussian states, for which the explicit solution can be found. Examples so far include state estimation \cite{GG}, teleportation benchmarks \cite{GBA}, state purification and dilution \cite{BGA} and quantum learning \cite{GKot}.

Suppose we are given $n$ independent $d$-dimensional quantum systems (qudits) each  prepared in the unknown
but mixed (full rank) state $\rho\in M(\mathbb{C}^d)$ with \emph{distinct eigenvalues}.  By means of an adaptive estimation strategy, we can effectively localize the initial state of each qudit within a neighborhood of size $n^{-\frac{1}{2}+\epsilon}$ centered at an estimate $\rho_0$. If  $\lambda_{1}>\dots>\lambda_{d}$ are the eigevalues of $\rho_{0}$ then with respect to its eigenbasis, any neighbouring state can be written as
\begin{equation}\label{matrice1ord}
\rho_{(\textbf{M}/\sqrt{n},\boldsymbol{\lambda})}=\rho_{0}+ \frac{1}{\sqrt{n}}\left(
\begin{array}{cccc}
u_1&\Lambda_{1,2}z^*_{1,2}&\hdots&\Lambda_{1,d}z^*_{1,d}\\
\Lambda_{1,2}z_{1,2}&u_2&\ddots&\vdots\\
\vdots&\ddots&\ddots&\Lambda_{d-1,d}z^*_{d-1,d}\\
\Lambda_{1,d}z_{1,d}&\hdots&\Lambda_{d-1,d}z_{d-1,d}&-\sum^{d-1}_{i=1}u_i
\end{array}\right)
\end{equation}
where $\textbf{M}=(\textbf{z},\textbf{u})\in  \mathbb{C}^{d(d-1)/2}\times \mathbb{R}^{d-1}$ is a displacement parameter, and $\Lambda_{jk}=\sqrt{(\lambda_j-\lambda_k)/2}$ are constant coefficients chosen for later convenience. We can then define the local quantum statistical model around $\rho_0$ as
\begin{equation}
\mathcal{Q}_n=\left\{\rho^{\otimes n}_{\big(\textbf{M}/\sqrt{n},\, \lambda\big)}:\|\textbf{M}\|\leq n^\epsilon\right\}.
\label{qmod}
\end{equation}
Let
\begin{equation}\label{eq.o.i}
\{O_1,....,O_{d^2-1}\}=\left\{q_{1,2},p_{1,2},....,q_{d-1,d},p_{d-1,d},b_1,....,b_{d-1}\right\}.
\end{equation}
be the selfadjoint operators in $M(\mathbb{C}^d)$ defined as
\begin{equation}\label{eq.basis}
q_{j,k} := \frac{|j\rangle \langle k| + |j\rangle \langle k|}{\sqrt{2(\lambda_{j} -\lambda_{k})}}, \quad
p_{j,k} :=  \frac{ i(  |k\rangle \langle j| - |j\rangle \langle k|)}{\sqrt{2(\lambda_{j} -\lambda_{k})}}, \quad
b_{i}:= |i\rangle\langle i| -\lambda_{i} \mathbf{1}.\quad
\end{equation}
One can verify easily that $O_{a}$ satisfy the following
properties:
\begin{enumerate}
\item
$\{O_1,....,O_{d^2-1}\}$ is a basis in the space of operators with $\text{Tr}(\rho_{0} O)=0$;

\item
$\text{Tr}(\rho_{0} [O_a,O_b]) =i\Omega_{a,b}$ where $\Omega$ is the $(d^2-1 ) \times (d^2-1)$ block diagonal
symplectic matrix $\Omega=\text{Diag}(\sigma, \dots ,\sigma,{\bf 0}_{d-1} ).$

\item
The covariance matrix  $\textbf{V}$ with
\begin{equation}\label{eq.variance}
V_{ab}:= \text{Tr}(\rho_{0} O_a\ast O_b)
\end{equation}
has all elements equal to zero except
\begin{equation}\label{eq.covariance.qudits}
\text{Tr}(\rho_{0} q_{j,k}^{2})= \text{Tr}(\rho_{0} p_{j,k}^{2}) =v_{j,k}:= \frac{\lambda_{j}+\lambda_{k}}{2(\lambda_{j} -\lambda_{k})} , \quad
\text{Tr}(\rho_{0} b_{j}b_{k}):= V^{\rm cl}_{jk} := \delta_{jk} \lambda_{j}  -\lambda_{j}\lambda_{k}.
\end{equation}
\end{enumerate}

%

Let $O_i(n)\in M(\mathbb{C}^{d})^{\otimes n}$ denote the corresponding collective (fluctuation) observables
\begin{equation}\label{osserv}
O_i(n) := \sum_{s=1}^{n} O_i^{(s)} ,\qquad  O_i^{(s)}:=
\mathbf{1}\otimes \dots \otimes O_i\otimes \dots \otimes \mathbf{1},
\end{equation}
with $O^{(s)}_i$ acting on the position $s$ of the tensor product. The collective observables play the role of sufficient statistics for our  model, and we would like to understand their asymptotic behaviour. Since all systems are independent and identically prepared, and the terms in each collective observable commute, we can apply classical Central Limit techniques to show that the following convergence in distribution holds for the collective states
$\rho^{\otimes n}_{\textbf{M}/\sqrt{n},\, \lambda}$,
\begin{eqnarray}
\frac{q_{j,k} (n) }{\sqrt{n}}                    &\overset{\mathcal{L}}{\longrightarrow} &
N\left(\mathrm{Re} (z_{j,k}), v_{j,k}\right),
\quad\qquad 1\leq j<k\leq d;
\nonumber\\
\frac{p_{j,k} (n) }{\sqrt{n}}                    &\overset{\mathcal{L}}{\longrightarrow} &
N\left(\mathrm{Im}(z_{j,k}), v_{j,k}\right),
\quad\qquad 1\leq j<k\leq d; \nonumber \\
\frac{b_{l} (n) }{\sqrt{n}}   &\overset{\mathcal{L}}{\longrightarrow} &
N\left(u_{l}, \lambda_{l}(1-\lambda_{l}) \right), \qquad
1\leq l \leq d-1\,, \nonumber
\end{eqnarray}
The key observation is that the unknown parameters $\bf {M}=(\bf{z},\bf{u})$ are recovered as means of the limit Gaussian distributions. However the limit model is \emph{not} a classical one due to the fact that the collective variables do not commute with each other. Therefore we need to take into account the commutation relations of the limit variables by using the quantum Central Limit Theorem (CLT). These
form a general continuous variable system as described in section \ref{GF}, with $d^{2}-1$ coordinates
${\bf Z}= (Z_1,\dots, Z_{d^2-1}) = (Q_{j,k}, P_{j,k}, B_{i})$  having the commutation relations
$$
[ Z_a, Z_b ] = {\rm Tr}(\rho_{0} [O_a, O_b ])\, \mathbf{1}. 
$$
By property (ii), this means that $(Q_{j,k}, P_{j,k})$ are canonical coordinates of $d(d-1)/2$ mutually commuting one-mode systems, and $B_{i}$ are classical random variables in the sense that they commute with each other and with the quantum coordinates $(Q_{j,k}, P_{j,k})$. By the CLT the limit state is the Gaussian
$\Phi(\bf{M}, \bf{V})$ with mean ${\bf M}$ and covariance matrix ${\bf V}$ defined in property (iii), equation
\eqref{eq.covariance.qudits}.
In particular, the individual modes $(Q_{j,k}, P_{j,k}, B_{i})$ are independent of each other and of the classical variables
$B_{i}$, and the latter have covariance matrix ${\bf V}^{\rm cl}$, cf. \eqref{eq.covariance.qudits}.


We define now the quantum-classical Gaussian statistical model $\mathcal{P}$ as
\begin{equation}
\mathcal{P}:= \{\Phi(\textbf{M},\textbf{V}) : \textbf{M}\in \mathbb{R}^{d-1}\times \mathbb{C}^{d(d-1)/2}\}.
\label{gmod}
\end{equation}
and enunciate the Local Asymptotic Normality Theorem \cite{GK2} which is used in establishing the optimality results in section \ref{c3}.
\begin{theorem}
\label{main}
The sequence of  qudit  models $\mathcal{Q}_n$ defined in \eqref{qmod} converges to the quantum-classical Gaussian model $\mathcal{P}$ defined in (\ref{gmod}), in the sense that there exist $\epsilon>0$ and channels
\begin{eqnarray}
T_n &:&
M(\mathbb{C} ^{d})^{\otimes n} \to \mathcal{T}_{1}
\left(L^{2}(\mathbb{R})^{\otimes d(d-1)/2}\right)\otimes  L^{1}(\mathbb{R}^{d-1}) ,
\nonumber\\
S_n &:&   \mathcal{T}_{1}
\left(L^{2}(\mathbb{R})^{\otimes d(d-1)/2}\right)\otimes L^{1}(\mathbb{R}^{d-1})\to M(\mathbb{C} ^{d})^{\otimes n},
\nonumber
\end{eqnarray}
such that
\begin{align}
\lim_{n\rightarrow\infty}\sup_{\|\textbf{M}\|\leq n^\epsilon}&
\left\|T_n(\rho^{\otimes n}_{(\textbf{M}/\sqrt{n},\lambda)})-\phi(\textbf{M},\bf{V})\right\|_1=0\nonumber\\
\lim_{n\rightarrow\infty}\sup_{\|\textbf{M}\|\leq n^\epsilon}&
\left\|\rho^{\otimes n}_{(\textbf{M}/\sqrt{n},\lambda)}-S_n(\phi\big(\textbf{M},\bf{V})\big)\right\|_1=0.
\end{align}
\end{theorem}
Note that the statement of the above theorem is different from that of the CLT in that Theorem \ref{main} shows that the collective state of the ensemble can be transferred by means of physical quantum channels to a Gaussian state, with vanishing norm-one error uniformly over the unknown parameters. The CLT is instead a statement about the convergence in law, for a fixed state and does not have an immediate operational interpretation. In the next section we will use LAN to transform the qudit reversal problem into a corresponding Gaussian reversal one, thus exploiting the solution to the latter obtained in section \ref{c2}.

\section{Optimal channel reversal on mixed qudit ensembles}
\label{c3}

We now focus on the main aim of this paper, that is to find optimal channels
$$
R_n:M(\mathbb{C}^d)^{\otimes n}\to M(\mathbb{C}^d)^{\otimes m}
$$
which reverse (at rate $\Gamma= m/n$) the action of the tensor product channel $C^{\otimes n}$  acting on $n$ identically prepared qudits, cf. \eqref{prob} [see Fig.~\ref{figproblem}(a)].  The performance of channel reversal can be quantified by a figure of merit, or risk, given by the trace-norm error
$
\left\|R_n(\rho^{\otimes n})-\rho^{\otimes m} \right\|_1.
$
We adopt a frequentist approach and look to minimize the maximum risk over all input states. We actually work with a more refined version of the maximum risk known as the {\it local} maximum risk, which was already employed in other quantum statistical problems \cite{GK,GJK,GKot,BGA}. For each state $\rho$, this is defined by
$$
\mathcal{R}_{\max}(R_n,\rho,\Gamma):=\sup_{\|\tau-\rho\|\leq n^{-\frac{1}{2}+\epsilon}}\left\|R_n(\tau^{\otimes n})-\tau^{\otimes m} \right\|_1
$$
and quantifies the worst performance of $R_{n}$ in a $n^{-\frac{1}{2}+\epsilon}$-neighbourhood of $\rho$.
This restriction does not amount to making an assumption about the location of the unknown state, since
using a small sample $n^{1-\epsilon}\ll n$ of the systems, one can effectively localize the unknown state within a confidence region of size $n^{-\frac{1}{2}+\epsilon}$.
We aim to find
an optimal reversing strategy whose asymptotic risk is equal to the local minimax risk
\begin{equation}\label{minirisiko}
\mathcal{R}_{\text{minmax}}(\rho,\Gamma):=\limsup_{n\rightarrow\infty}\inf_{R_n}\mathcal{R}_{\max}(R_n,\rho,\Gamma)\,.
\end{equation}

The problem of finding the optimal reversing channel is thus reformulated as follows. For a given state $\rho$ and output rate $\Gamma=m/n$, what is the minimax risk $\mathcal{R}_{\text{minmax}}(\rho, \Gamma)$ and which procedure achieves it? We will not solve this problem for all
$(\rho, \Gamma)$, but rather we will show that for each $\rho$ there is an interval $[0,\Gamma^{\star}]$ for which the reversal can be performed perfectly (i.e. with asymptotically vanishing risk $\mathcal{R}_{\text{minmax}}(\rho,\Gamma)=0$), and we will subsequently identify an asymptotically optimal sequence of reversing channels $R^\star_n$ which achieves the maximum rate $\Gamma^{\star}$.

As anticipated, the quantum LAN theory \cite{GK, GJK, GK2, GJ, GKJ}, reviewed in section \ref{LF} is the key tool in  solving this problem. For technical reasons related to the validity of LAN, we assume that the state $\rho$ is not on the boundary of the state space, and is not degenerate so that its spectrum $\boldsymbol{\lambda}= (\lambda_{1},\dots ,\lambda_{d})$ satisfies $\lambda_{1}> \dots >\lambda_{d}$. A typical random state satisfies these assumptions. Following equations \eqref{matrice1ord}-\eqref{eq.basis} we can parametrize the states in the $n^{-1/2+\epsilon}$ neighbourhood of $\rho$ as $\rho_{\bf{M}/\sqrt{n},\,\boldsymbol{\lambda}}$ where $\textbf{M}=({\bf z},{\bf u})\in\mathbb{C}^{d(d-1)/2}\times  \mathbb{R}^{d-1}$ is a vector of expectations for the matrices $\{O_{1},\dots , O_{d^{2}-1} \}$. By LAN, the sequence of input quantum statistical models
 $$
 \mathcal{Q}_{n}:= \left\{ \rho^{\otimes n}_{(\textbf{M}/\sqrt{n},\boldsymbol{\lambda})} : \|{\bf M}\|\leq n^{\epsilon}  \right\}
 $$
 is asymptotically equivalent in the sense of Theorem~\ref{main}, to the sequence of Gaussian models $\mathcal{P}_{n}$ over a quantum-classical continuous variable system with coordinates ${\bf Z}$
 $$
\mathcal{P}_{n}:= \left\{ \Phi({\bf M}, {\bf V})
:  \|{\bf M}\|\leq n^{\epsilon}\right\}
$$
where $ \Phi({\bf M}, {\bf V})$ which is a tensor product between  $d(d-1)/2$ independent one-mode displaced  thermal equilibrium states $ \Phi(z_{j,k},v_{j,k})$ (one for each pair $j<k$) with mean $z_{j,k}$ and covariance $v_{j,k}$ (cf. equation
\eqref{eq.covariance.qudits}), and a classical $(d-1)$-dimensional Gaussian probability density $\mathcal{N}({\bf u}, {\bf V}^{\rm cl})$.

We apply the same procedure to the output state $\rho^{\prime}:=C(\rho)$ and denote by
$$
\rho_{{\bf M}^{\prime}/\sqrt{n},\boldsymbol{\lambda}^{\prime}}:= C(\rho_{{\bf M}/\sqrt{n},\boldsymbol{\lambda}})
$$
the transformed state belonging to the neighbourhood of $C(\rho)$. The corresponding Gaussian model with coordinates
${\bf Z}^{\prime}$ is
$$
 \mathcal{P}^{\prime}_{n}:= \left\{ \Phi({\bf M}^{\prime}, {\bf V}^{\prime})
:  \|{\bf M}^{\prime}\|\leq n^{\epsilon}\right\}
 $$
where ${\bf V}^{\prime}$ depends solely on the spectrum of $\rho^{\prime}$. The displacements of the two Gaussian models are related by a linear transformation
$$
X^{T}:{\bf M}\mapsto {\bf M}^{\prime}
$$
which describes the local action of $C$ around $\rho$, and hence depends only on these two objects. The ensemble transformation $C^{\otimes n}$ acts on the localized states as
\begin{equation}
C^{\otimes n}:\rho^{\otimes n}_{(\textbf{M}/\sqrt{n},\boldsymbol{\lambda})}\longmapsto\rho^{\otimes n}_{(\textbf{M}'/\sqrt{n},\boldsymbol{\lambda}')}.
\label{gqch}
\end{equation}
Since both the input and the output and can be approximated by Gaussian states and the transformation between the displacements is linear, one can expect that the two Gaussian states are connected by a Gaussian channel $G$ as depicted in diagram (\ref{cd}). The following theorem states that this is indeed the case and the proof can be found in  \ref{proofexists}.
\begin{theorem}\label{teoG}
There exists a Gaussian channel $G$
such that
$$
G:\phi({\bf M}, {\bf V}) \mapsto \phi({\bf X}^{T}{\bf M}, {\bf V}^{\prime}).
$$
The matrices $({\bf X}_{G},{\bf Y}_{G})$ characterizing $G$ (cf. \eqref{gch}) are ${\bf X}_{G}={\bf X}$ and ${\bf Y}_{G}= {\bf V}^{\prime} - {\bf X}^{T}{\bf V}{\bf X}$.
\end{theorem}

We now move to the problem of finding the optimal reversing channel for multiple qudits. As illustrated in the right side of the diagram (\ref{cd}), the solution relies on LAN to recast the problem into the Gaussian one solved in Theorem \ref{teogauss}. We now state the second main result of this paper, whose proof is in \ref{secproofqudits} .
\begin{theorem}\label{teoqudits}
Let   $\rho\in M(\mathbb{C}^{d}) $ be a full rank state with non-degenerate spectrum, and let
$C: M(\mathbb{C}^{d})\to M(\mathbb{C}^{d})$ be a quantum channel which is invertible as a linear map.
Let $(k^{\star}, H^{\star})$ be the optimal rate and channel for the Gaussian reversal problem
$(cf. ~Theorem~\ref{teogauss})$
$$
\phi(\textbf{M},{\bf V})\stackrel{G}{\longmapsto}\phi({\bf X}^{T}\textbf{M},{\bf V}')\stackrel{H}{\longmapsto}\phi(k\textbf{M},{\bf V})\, \qquad
{\bf M}\in \mathbb{R}^{d-1}\times \mathbb{C}^{d(d-1)/2}
$$
where ${\bf V}, {\bf V}^{\prime}, X$ depend on $C$ and $\rho$ as above.

Then the maximum reversal rate for the qudit ensemble is $\Gamma^{\star}=(k^{\star})^2$ and
$$
R^\star_n:=S_{\Gamma^{\star} n}\circ H^\star\circ T_n\,,
$$
is a sequence of optimal reversing channels i.e.
$$
\limsup_{n\rightarrow\infty} \sup_{ \|\tau-\rho\|_{1} \leq n^{-\frac{1}{2}+\epsilon}}
\left\| R_n(\tau^{\otimes n}) - \tau^{\otimes m} \right\|_1 =0.
$$

\end{theorem}

This result shows, in full generality, that in order to reverse the action of an arbitrary channel acting on a large ensemble of $n$ qudits, independent and identically prepared in an unknown mixed state $\rho$, the optimal strategy is to take a shortcut through phase space \cite{noteshortcut}, optimally reverse the corresponding Gaussian channel, and then map the output back onto $m$ qudits (with rate $m/n=\Gamma^{\star}$). 

This kind of digital-analogue interconversion might prove handy when it comes to counteract the effect of noisy channels affecting quantum states of, say, $n$-atom ensembles: The best strategy ought to be transferring their state onto light modes (e.g. using quantum non-demolition interactions \cite{polzik}), implementing corrective procedures on the obtained Gaussian beams in phase space (according to the recipe of Theorem~\ref{teogauss}), and then mapping the restored state back onto the atomic storage unit.
We will now showcase our results on a specific example.

\section{Example: reversal of the phase flip qubit channel}
\label{c4}

The qubit phase flip channel  
is defined by $C(\rho)=\sum_k E_k\rho E^\dagger_k$,
with $E_0=\sqrt{p}\ \text{Diag}(1,1)$ and $E_1=\sqrt{1-p}\  \text{Diag}(1,-1)$ \cite{NC}. Phase flip is a common source of error in quantum computation and several algorithms have been developed for its correction \cite{errorcorrection}. According to Theorem \ref{teoqudits}, an optimal way to reverse the composite action of individual phase flips on a $n$-qubit register is to map the state of the qubit register onto a Gaussian state (which in this case comprises a single quantum mode and one real classical random variable), reverse the corresponding Gaussian channel (up to a rescaling factor $k^{\star}$), and map the output back onto the qubits.

\begin{figure}[tb]
\begin{center}
 \includegraphics[width=5.5cm]{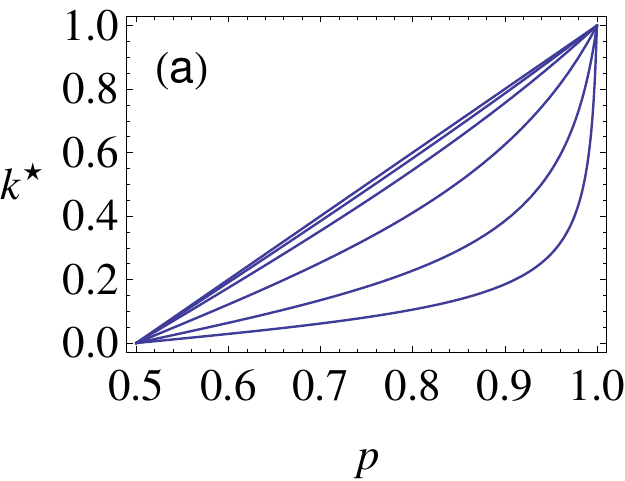}
\hspace*{0.5cm}
\includegraphics[width=5.5cm]{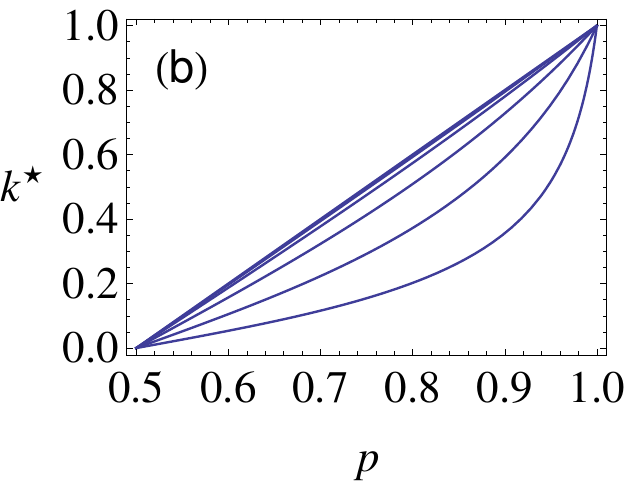}
\caption{(Color online) Plots of $k^{\star}$ as a function of $p$ for different values of $\|\textbf{r}\|$. From top to bottom in both panels: $\|\textbf{r}\| = 0, 0.3, 0.5, 0.8, 0.95, 0.99$. In panel (a), $r_z=0$; in panel (b), $r_z=2r_x$.}
\label{figk0vsp}
\end{center}
\end{figure}

As described in the previous section we consider states in a neighbourhood of $\rho$; the latter has in general the Bloch vector representation
\begin{equation}
\rho=\frac{1}{2}\left(\begin{array}{cc}
1+r_z&r_x+ir_y\\
r_x-ir_y& 1-r_z
\end{array}\right)
\label{state}.
\end{equation}
The neighbourhood of $\rho$ is parametrized as in \eqref{matrice1ord} with ${\bf M}=(z,u)\in \mathbb{C}\times  \mathbb{R}$.
The action of the channel $ C$ is
$$
C:{\bf r} =(r_x,r_y,r_z)\quad\mapsto\quad{\bf r}^{\prime}=\big((2p-1)r_x,(2p-1)r_y,r_z\big).
$$
From this we can compute the covariance matrices
$$
{\bf V}=\text{Diag}\left(\frac{1}{2\|\textbf{r}\|},\frac{1}{2\|\textbf{r}\|},1-\|\textbf{r}\|^2\right),
\quad {\bf V}'=\text{Diag}\left(\frac{1}{2 \|{\bf r}^{\prime}\|},\frac{1}{2 \|{\bf r}^{\prime}\|},1-  \|{\bf r}^{\prime}\|^2\right),
$$
where $\|{\bf r}^{\prime}\|^{2}=(2p-1)^2(r^2_x+r^2_y)+r^2_z$. Similarly the matrix ${\bf X}_G^T$ is obtained after computing the transformation of the local states under $C$
\begin{equation}
{\bf X}_G^T =
\left(
\begin{array}{ccc}
 (2 p-1) \sqrt{\frac{\|\textbf{r}\|}{ \|{\bf r}^{\prime}\|}} & 0 & 0 \\
 0 & (2 p-1) \big(\frac{\|\textbf{r}\|}{ \|{\bf r}^{\prime}\|}\big)^{\frac{3}{2}} & 0 \\
 0 & -\frac{4 (1-p) p r_x r_z}{ \|{\bf r}^{\prime}\|\cdot \|\textbf{r}\|} & \frac{ \|{\bf r}^{\prime}\|}{\|\textbf{r}\|}
\end{array}
\right)
\,.
\label{gflip}
\end{equation}


We are interested in evaluating the threshold factor $k^{\star}$ as given by (\ref{maxk}), which is associated to the optimal rate
$m/n=(k^{\star})^2$ for phase flip reversal on the qubit register.
Due to the inherent symmetry of the channel, we may set $r_y=0$ without any loss of generality, and assume $\frac12 \leq p \leq 1$. Therefore $k^{\star}$ will depend on $p, r_x, r_z$. The results are summarized as follows.

\begin{figure}[t]
\begin{center}
\includegraphics[width=9cm]{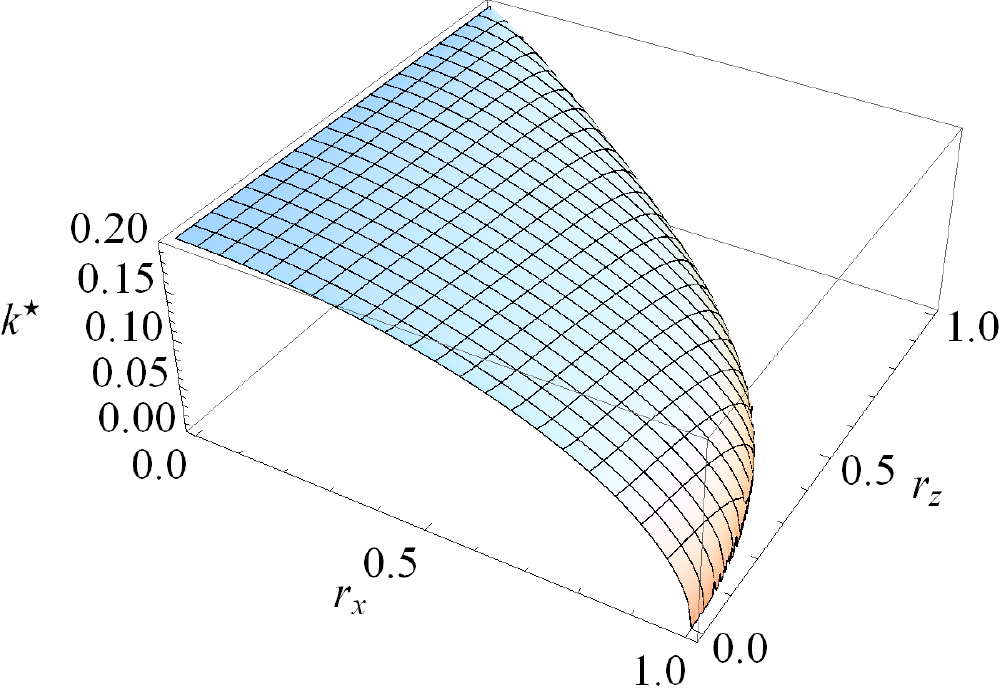}
\caption{(Color online) Plot of $k^{\star}$ as a function of $r_x$ and $r_z$ for $p=0.6$.}
\label{figk0vsrxry}
\end{center}
\end{figure}

For $p=1$ the channel is trivially the identity, therefore  $k^{\star}=1$ for any $\textbf{r}$. For  $p=1/2$ the channel is never reversible, and $k^{\star}=0$ for any $\textbf{r}$. In the limit of pure input states of each qubit ($|\textbf{r}|\rightarrow 1$), we find that that $k^{\star} \rightarrow 2p-1$ if $r_x\rightarrow 0$, and $k^{\star} \rightarrow 0$ otherwise (i.e., the channel is not reversible for any pure state, if there is a nonzero component on the $x$ axis). In case $r_x=0$,  we find $k^{\star}(p,r_x=0,r_z)=2p-1$, which means that, as intuitively expected, the rate does not depend on $r_z$. On the other hand, in case  $r_z=0$, we have
$$
k^{\star}(p,r_x,r_z=0)= \sqrt{[(2p-1)^2(1-r_x^2)]/[1-(-1+2 p)^2 r_x^2]}.
$$
We observe that for small values of $r_x$ this tends again to the line $2p-1$; for $r_x$ close to $1$, $k^{\star}$ is instead  a sublinear function of $p$, as depicted in Fig.~\ref{figk0vsp}, showing that the reversal becomes inefficient in terms of the number of perfectly retrieved copies.
In general, if a nonzero $r_x$ is present as well, the behavior of $k^{\star}$ as a function of $p$ is qualitatively similar to the previous case, but typically $k^{\star}$ increases, at fixed $\|\textbf{r}\|$ and $p$, with increasing ratio $|r_z/r_x|$. The general shape of $k^{\star}$ as a function of the Bloch vector components interpolates among the various limits discussed above, and is plotted in Fig.~\ref{figk0vsrxry}.

\section{Conclusions}\label{secconcl}
In this paper we have solved the general problem of optimal channel reversal for an ensemble of generally mixed independent and identically prepared qudits. This substantially extends our earlier work \cite{BGA} on the particular instance of state purification (and dilution) of qubit ensembles. To accomplish our task, we have employed the versatile statistical tool of LAN \cite{GK,GK2} to recast the problem in terms of Gaussian states and channels, then solved the  open problem of optimal reversal of a general Gaussian channel acting on a quantum-classical Gaussian state with  a given covariance.


The methods demonstrated in this paper provide powerful strategies for counteracting undesired noise effects in quantum memories and long-distance quantum communications, based on interfaces between discrete ensembles and continuous modes \cite{atomicmemories,polzik,quantuminternet}.
A further generalization of our work, worth addressing in the future, could be that of considering the optimal reversal of quantum channels acting on {\it correlated} copies of $n$ qudits, possibly requiring an extension of the LAN theory beyond the paradigm of independent and identically prepared systems.

\ack

We thank M. Mosonyi and N. Datta for discussions. We acknowledge the University of Nottingham funding through ECRKTA/2011 and the Additional Sponsorship grant RDF/BtG/0612b/31. MG was supported by the EPSRC Fellowship EP/E052290/1.

\appendix

\section[$ $\qquad  \qquad Proof that the Gaussian Channel $G$ in the diagram (\ref{cd}) exists]{Proof that the Gaussian Channel $G$ in the diagram (\ref{cd}) exists}
\label{proofexists}

Note that by the LAN construction the coordinates ${\bf Z}$ and
${\bf Z}^{\prime}$  of the input and output Gaussian continuous variable systems have the same symplectic matrix $\Omega$,
\begin{equation}
\label{eq.syplectic.equality}
[Z_{a} , Z_{b}] =   \text{Tr}( \rho [O_{a}, O_{b}]) \mathbf{1}= i \Omega_{a,b} \mathbf{1}=  \text{Tr}( \rho [O_{a}^{\prime}, O_{b}^{\prime}])=
[Z_{a}^{\prime} , Z_{b}^{\prime}]
\end{equation}
and the covariance matrices are determined by the states $\rho$ and $\rho^{\prime}=C(\rho)$
\begin{equation}
\label{eq}
\langle Z_a \ast Z_b\rangle = \text{Tr}( \rho O_{a}\ast O_{b}), \qquad
\langle Z_a^{\prime} \ast Z_b^{\prime}\rangle = \text{Tr}(\rho^{\prime} O_{a}^{\prime}\ast O_{b}^{\prime}).
\end{equation}

 Recall that any Gaussian channel from the input to the output is of the form
$$
G^{*}: W_{z}\longmapsto W_{{\bf X}_G z }e^{-\frac{1}{2}z^T{\bf Y}_Gz}
$$
where  ${\bf X}_G$, ${\bf Y}_G$ are  real matrices of dimension $d^{2}-1$,
with ${\bf Y}_{G}$ positive and satisfying the matrix inequality
\begin{equation}\label{eq.yg}
{\bf Y}_G\geq  i({\bf X}_G^T\Omega {\bf X}_G-\Omega).
\end{equation}
Since the means are transformed as ${\bf M}\mapsto {\bf X}_{G}^{T}{\bf M}$ the matrix ${\bf X}_{G}$ must be equal to
${\bf X}$. Moreover the output variance is
$$
{\bf V}^{\prime} = {\bf X}^T {\bf V} {\bf X} + {\bf Y}_{G} 
$$
which determines the second matrix  ${\bf Y}_{G}= {\bf V}^{\prime}- {\bf X}^T {\bf V} {\bf X}$. It remains to show that ${\bf Y}_{G}$ satisfies \eqref{eq.yg} which is equivalent to
\begin{equation}
{\bf V}^{\prime} + i\Omega \geq {\bf X}^T({\bf V}+i\Omega){\bf X}.
\label{in}
\end{equation}
By using \eqref{eq.syplectic.equality} and \eqref{eq} the latter can be translated into the following property
of $C$
\begin{equation}
\widetilde{\textbf{V}}'\geq {\bf X}^T\widetilde{\textbf{V}} {\bf X}
\label{cp}
\end{equation}
where $\widetilde{\textbf{V}}'_{a,b}=\text{tr}\big(C(\rho) O'_a O'_b\big)$ and
$\widetilde{\textbf{V}}_{a,b}=\text{tr}(\rho O_a O_b)$.
Finally, to verify this inequality note that for all ${\bf M}$
$$
 \text{Tr}\left(\rho_{(\textbf{M},\boldsymbol{\lambda})}{{\bf X}^T}{\bf O}\right)= {\bf X}_G^T\textbf{M}=
\text{Tr}\left(C(\rho_{(\textbf{M},\boldsymbol{\lambda})}) {\bf O}'\right)
=\text{tr}\big(\rho_{(\textbf{M},\boldsymbol{\lambda})}C({\bf  O}')\big)
$$
which implies ${\bf O}'={{\bf X}^T}{\bf O}$. This result together with the inequality $T(A^*A)\geq T(A^*)T(A)$ valid for any channel $T$ imply \eqref{cp}:
\begin{align}
{\bf c}^{\dagger} \widetilde{\bf V}^{\prime}{\bf c}= \sum_{a,b} c_{a}^{*} c_{b}
\text{Tr}(C(\rho)  O'_a O'_b)&=
\sum_{a,b} c_{a}^{*} c_{b}
\text{Tr}\big(\rho  C^{*}(O'_a O'_b)\big)\nonumber\\
&\geq
\sum_{a,b} c_{a}^{*} c_{b}
\text{Tr}\big(\rho C(O'_a)C(O'_b)\big)\nonumber\\
&=
{\bf c}^{\dagger} {\bf X}^{T}\widetilde{\bf V} {\bf X}{\bf c}.
\end{align}
Hence if $C$ is a completely positive map then $G$ is a Gaussian channel.
\hfill $\square$

\section[$ $\qquad  \qquad Proof of Theorem \ref{teoqudits}]{Proof of Theorem \ref{teoqudits}}
\label{secproofqudits}

We need to show that the reversal of $C$ is possible on the interval $[0, \Gamma^{\star}]$ and not possible for
$\Gamma>\Gamma^{\star}$.
Beginning with the qudit state
$$
C(\rho_{(\textbf{M}/\sqrt{n},\, \boldsymbol{\lambda})})^{\otimes n}  =\rho^{\otimes n}_{(\textbf{M}'/\sqrt{n},\, \boldsymbol{\lambda}')},
$$ where $\textbf{M}'={\bf X}_G^T\textbf{M}$, we apply the LAN channel $T_n$ (cf. Theorem \ref{main}) followed by $H^\star$ (cf. Theorem \ref{teogauss}) to obtain a state which is asymptotically undistinguishable from the Gaussian state $\Phi(k^{\star}\textbf{M},{\bf V})$. To this we apply the inverse LAN map $S_m=S_{k^{\star 2} n}$ to achieve the corrected state
$$
R_{n}\left(\rho^{\otimes n}_{(\textbf{M}'/\sqrt{n},\, \boldsymbol{\lambda}')}\right) = S_{m} \circ H^{\star} \circ T_{n} \circ C^{\otimes n} \left(\rho_{(\textbf{M}/\sqrt{n},\, \boldsymbol{\lambda}) }\right).
$$
We first show that ${R}^\star_n$ is an (asymptotic) reversing channel for $\rho^{\otimes n}$:
 \begin{align*}
 \mathcal{R}_{\bf max}(R^\star_n,\rho,\Gamma)&=
\sup_{\| {\bf M}\| \leq n^{\epsilon}} \left\|\rho^{\otimes m}_{(\textbf{M}/\sqrt{n},\boldsymbol{\lambda})} -
{R}^\star_n\left(\rho^{\otimes n}_{(\textbf{M}'/\sqrt{n},\boldsymbol{\lambda}')}\right)\right\|_1\nonumber\\
 &\leq
\sup_{\| {\bf M}\| \leq n^{\epsilon}} \left\|\rho^{\otimes m}_{(\textbf{M}/\sqrt{m},\boldsymbol{\lambda})}-S_m\left(\phi( k^{\star}\textbf{M},{\bf V})\right)\right\|_1
 \nonumber \\ &\quad +
\sup_{\| {\bf M}\| \leq n^{\epsilon}} \left\|S_m\left(\phi(k^{\star}\textbf{M},{\bf V})\right)- S_m\circ H^\star\circ T_n(\rho^{\otimes n}_{(\textbf{M}',\boldsymbol{\lambda}')})\right\|_1
 \nonumber\\
&\leq
\sup_{\| {\bf M}\| \leq n^{\epsilon}} \left\|\phi(k^{\star}\textbf{M},{\bf V})- H^\star\circ T_n\left(\rho^{\otimes n}_{(\textbf{M}',\boldsymbol{\lambda}')}\right)\right\|_1+o(1)\nonumber\\
&=
 \left\|\Phi(k^{\star}\textbf{M},{\bf V})-H^\star\left(\phi(\textbf{M}',{\bf V}')\right)\right\|_1+o(1)=o(1),\nonumber\\
 \end{align*}
where we have used the contractivity of the trace norm under quantum operations, the properties of LAN and the fact that
$H^{\star}$ is a Gaussian reversing channel for $G$, cf. \eqref{minmax}.

We now show that it is impossible to obtain asymptotically exact reversal for  $\Gamma>\Gamma^{\star}$. Indeed, suppose there exists a sequence of reversing channels
$\tilde{R}_n: M(\mathbb{C}^{d})^{\otimes n}\to M(\mathbb{C}^{d})^{\otimes \Gamma n} $  such that
$$
\underset{n\to\infty}{\rm limsup} \sup_{\|{\bf M} \|\leq n^{\epsilon}}
\left\|\tilde{R}_n\left(\rho^{\otimes n}_{(\textbf{M}',\boldsymbol{\lambda}')}\right) -\rho^{\otimes m}_{(k\textbf{M},\boldsymbol{\lambda})} \right\|_1 =0.
$$
We then use LAN again to show that there exists a $G$-reversing channel $T_m\circ\tilde{R}_n\circ S_n$
with rate $k>k^{\star}$, which is impossible:
\begin{align*}
&\left\| T_m\circ\tilde{R}_n\circ S_n \left(\phi(\textbf{M}',{\bf V}'\right) -\phi\left(k\textbf{M},{\bf V} \right)\right\|_1\nonumber\\
&\leq \left\|\tilde{R}_n\left(\rho^{\otimes n}_{(\textbf{M}',\boldsymbol{\lambda}')}\right) -\rho^{\otimes m}_{(k\textbf{M},\boldsymbol{\lambda})} \right\|_1  +
\left\| S\left(\phi(\textbf{M}',{\bf V}' )\right) -\rho^{\otimes m}_{(k\textbf{M},\boldsymbol{\lambda})} \right\|_1 +
\left\|T_m\left(\rho^{\otimes m}_{(k\textbf{M},\boldsymbol{\lambda})}\right)-
\phi(k\textbf{M},{\bf V})\right\|_1\nonumber\\
&= o(1)\,.
\end{align*}
\hfill
$\square$

\section*{References}
\bibliographystyle{iopart-num-j}

%
%
%
%
%
%
%

\end{document}